\documentclass[iop,apj,useAMS,usenatbib]{emulateapj-rtx4}

\usepackage{graphicx,epstopdf,epsfig}
\usepackage{lscape}
\usepackage{amsmath}
\usepackage{longtable}
\usepackage{natbib}
\usepackage[usenames,dvipsnames]{color}
\usepackage{gensymb}
\usepackage{multirow}
\usepackage{capt-of}
\usepackage{float}
\usepackage{amssymb} 
\usepackage[utf8]{inputenc} 
\usepackage{booktabs}
\usepackage{wasysym}
\usepackage{hyperref,url}
\usepackage{amsmath}
\usepackage{float}
\usepackage[export]{adjustbox}
\usepackage[caption = false]{subfig}
\usepackage{stackengine}
\hypersetup{colorlinks=true, citecolor=blue, filecolor=magenta, urlcolor=cyan, linkcolor=blue}

\begin{document}


\title{Photometric Redshift Estimation with Galaxy Morphology using Self-Organizing Maps}
\author{Derek Wilson\altaffilmark{1}, Hooshang Nayyeri\altaffilmark{1}, Asantha Cooray\altaffilmark{1}, Boris H\"{a}u{\ss}ler\altaffilmark{2}}

\altaffiltext{1}{Department of Physics \& Astronomy, University of California, Irvine, CA 92697, USA}
\altaffiltext{2}{European Southern Observatory, Alonso de Cordova 3107, Vitacura, Casilla 19001, Santiago, Chile}


\begin{abstract}

We use multi-band optical and near-infrared photometric observations of galaxies in the Cosmic Assembly Near-Infrared Deep Extragalactic Legacy Survey (CANDELS) to predict photometric redshifts using artificial neural networks. The multi-band observations span over 0.39\,$\mu$m to 8.0\,$\mu$m for a sample of $\sim$1000 galaxies in the GOODS-S field for which robust size measurements are available from {\it Hubble} Space Telescope Wide Field Camera 3 observations. We use Self Organizing Maps (SOMs) to map the multi dimensional photometric and galaxy size observations while taking advantage of existing spectroscopic redshifts at $0<z<2$ for independent training and testing sets. We show that use of photometric and morphological data led to redshift estimates comparable to redshift measurements from SED modeling and from self-organizing maps without morphological measurements.
  
\end{abstract}



\section{Introduction}

Photometric redshift (photo-z) estimation is crucial for astrophysical applications as obtaining spectroscopic redshifts for large samples of distant galaxies is often times infeasible. Physical properties of extragalactic sources further depend on accurate redshift measurements. The photometric redshift can also be used as a good proxy for distance for mapping the large scale structure and performing weak lensing studies (\citealp{Munshi2008}).

Unfortunately, due to selective sampling of the galaxy SED, photometric redshifts suffer from much higher uncertainties than spectroscopic redshifts. Errors in photometric redshifts can significantly affect cosmological parameter measurements in, for example, weak lensing studies (e.g., \citealp{Huterer2006, Ma2006, Bernstein2010}) and baryon acoustic oscillation studies (e.g., \citealp{Zhan2006, ChavesMontero2018}).

The observable quantity available for photo-z estimation is galaxy photometry in multiple wavelength bands, and a large number of techniques have been developed to estimate redshift while trying to minimize $z_{\rm phot} - z_{\rm spec}$. Photometric redshift estimation is primarily done via template fitting (e.g., \citealp{Lanzetta1996, FernandezSoto1999}) and/or statistical (e.g., \citealp{Connolly1995}) and machine learning techniques. As surveys grow ever larger, machine learning techniques that can process enormous amounts of data with minimal human input are becoming increasingly important.

Some techniques for photo-$z$ estimation involve using artificial neural networks with photometry and/or morphology data (e.g., \citealp{Firth2003, Ball2004, CollisterLahav2004, Vanzella2004, Bonfield2010, Soo2018}), support vector machines (e.g., \citealp{Wadadekar2005, JonesSingal2017}), the Multi-Layer Perceptron with Quasi Newton Algorithm ({\sc mlpqna}, \citealp{Brescia2013}), and the conditional density estimator {\sc flexcode} (\citealp{Izbicki2017}). Statistical models have also been developed, such as the surface brightness and photometry model of \citet{Kurtz2007}, the algorithm based on surface brightness, S\`ersic index and photometry developed in \citet{WrayGunn2008}, and the Gaussian process regression models (\citealp{WaySrivastava2006, Way2011, Bonfield2010, Almosallam2016a, Almosallam2016b}), which also appears in \citet{Gomes2018} when applied to infrared- and visible-band photometry in conjunction with angular size. \citet{Wadadekar2005} use support vector machines to estimate redshifts from photometric data as well as the $50\%$ and $90\%$ Petrosian radii for their sources. They observe a $15\%$ increase in accuracy when they use the two Petrosian radii with photometry than when photometry alone was used. The empirical techniques in \citet{VinceCsabai2006} use photometry and morphological data from SDSS, and they find that the weak correlation between morphology and redshift leads to only negligible gains in photo-z estimation accuracy. \citet{Singal2011} use a principal component analysis including morphological parameters to estimate photometric redshifts for the All-wavelength Extended Groth Strip International Survey (AEGIS; \citealp{Davis2007}). They conclude that the additional noise added to the data set by including morphological parameters will offset any of the gains coming from correlations between redshift and morphology. \citet{JonesSingal2017} use a support vector machine to estimate photometric redshifts. Their work includes principal components of eight morphological parameters; however, they observe no significant decrease in the RMS error or in the number of outliers (i.e., the number of galaxies with $(z_{\rm phot} - z_{\rm spec}) / (1 + z_{\rm spec})$ greater than some value, such as the value of 0.15 in \citet{Hildebrandt2010}) when using morphological data. Machine learning models are trained on photometric and/or morphological features that have been derived from the galaxy images. \citet{Hoyle2016} develops a deep neural network that is trained directly on galaxy images, so the network itself decides which parts of the image are important. The paper does not note a significant improvement in redshift accuracy. A similar approach is found in \citet{Menou2018}, which uses a multi-layer perceptron/convolutional neural network (MLP-convnet) architecture that analyzes galaxy-integrated features such as fluxes and colors using the MLP framework while adding in morphological information found by analyzing images directly with the convnet framework. They find that the MLP-convnet architecture does lead to a significant improvement in accuracy but has no effect on the number of outliers.

We now focus on the use of a machine learning technique known as a self-organizing map (SOM; \citealp{Kohonen1982, Kohonen1990}) has increased in the last decade. An SOM is an artificial neural network whose main advantage is its ability to reduce the dimensionality of input data while preserving the relationships between data points, thus making those relationships easier to visualize. We use the SOM to characterize the multi-dimensional space of observed galaxy Spectral Energy Distributions (SEDs). In the literature, \citet{Tagliaferri2003} combine multilayer perceptrons with self-organizing maps to analyze photometric data from SDSS. There is also {\sc mlz} (Machine Learning and photo-$z$, \citealp{CarrascoKindBrunner2013, CarrascoKindBrunner2014}) which performs two regression algorithms for computing photo-zs: {\sc tpz}, which uses prediction trees and random forests, and {\sc somz}, which uses self-organizing maps. SOMs are also used by \cite{Masters2015} to estimate redshifts and identify regions in galaxy color space where spectroscopic redshifts have not been obtained in past surveys. If these gaps could be filled in by future surveys, such a complete training set would be a powerful tool for photo-z estimation using machine learning.  Recent work by \citet{SpeagleEisenstein2017a} develops a photo-z technique that combines template-fitting methods with self-organizing maps. When trained on mock LSST and {\it Euclid} data, they find that their technique can predict redshifts to the accuracy required for {\it Euclid} weak lensing measurements (\citealp{SpeagleEisenstein2017a, SpeagleEisenstein2017b}).

In this paper, we explore the effect that the addition of galaxy morphology to SOM training data has on redshift estimation accuracy. This paper is organized as follows: Section 2 describes the catalog data from GOODS-S used in our study. In Section 3, we summarize the self-organizing map algorithm. Sections 4 and 5 discuss the performance of the self-organizing maps when photometry alone and photometry plus morphology, respectively, are used for training. The AB magnitude system is used, and a flat-$\Lambda$CDM cosmology of $\Omega_{m_0}$ = 0.27, $\Omega_{\Lambda_0}$ = 0.73, and $H_0$ = 70 $\rm{km} ~ \rm{s^{-1}} ~  \rm{Mpc^{-1}}$ is assumed. The code developed herein will be made publicly available at \url{https://github.com/derkwilson/PhotSOM}.


\section{Data}

\begin{figure}[!th]
  \centering
    \includegraphics[trim=0cm 0cm 0cm 0cm, scale=0.4]{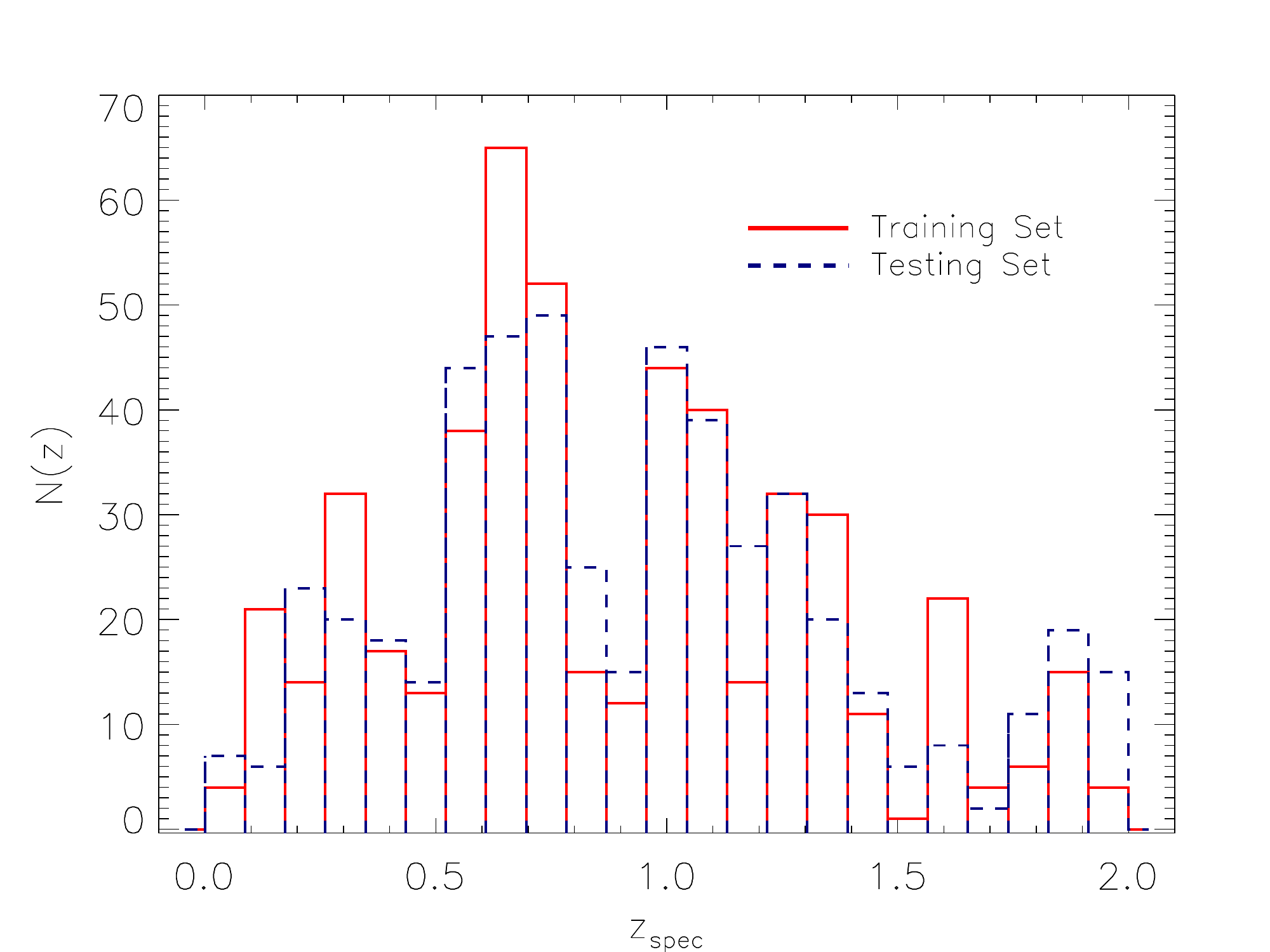}
  \caption{Histograms of the galaxy spectroscopic redshifts comprising the training (red) and testing (blue dashed) sets. The training and testing sets each contain 506 individual galaxies up to a redshift of 2.}
\label{fig:redshift_histograms}
\end{figure}

We use publicly available data from the GOODS-S field (centered at R.A. = 03$^{\rm{h}}$32$^{\rm{m}}$30$^{\rm{s}}$, Decl. = -27$^{\rm{d}}$48$^{\rm{m}}$20$^{\rm{s}}$) which covers an area of approximately 150 arcmin$^2$. Our training and testing catalogs are pulled from the Cosmic Assembly Near-Infrared Deep Extragalactic Legacy Survey (CANDELS; \citealp{Grogin2011, Koekemoer2011})\setcounter{footnote}{0}\footnote{\url{https://archive.stsci.edu/prepds/candels/}}. The full CANDELS GOODS-S catalog (\citealp{Guo2013}) includes optical, near-, and mid-infrared photometry from the {\it Hubble} Space Telescope (HST), the Very Large Telescope (VLT), and the Spitzer Infrared Array Camera (IRAC). Our primary training and testing catalogs each consist of 506 galaxies in the GOODS-S field with colors computed from the 15 bands listed in Table 1, comparable to the training and testing sets of \citet{Dahlen2013}. We have an additional training set with about 1360 sources, and the results using this training set do not differ significantly from the 506-source training set, so we will focus on the results from the 506-source set. We note that \citet{Bonfield2010} find that photo-z estimates deteriorate with fewer than 2000 training objects when using artificial neural networks and Gaussian process regression, but that the size and architecture of the network may permit reasonable results with fewer training objects. All sources in the training and testing sets have $z_{spec}<2$, and the distribution of redshifts is shown in Figure \ref{fig:redshift_histograms}. \citet{Dahlen2013} previously released a training/testing catalog set with photometry in the same bands (except ACS F814W) extending up to $z \sim 5$ in redshift, so we also test our SOMs with these catalogs for comparison.


In addition to the photometry, we use half-light radii (\citealp{Haussler2013}) and concentration, asymmetry, and smoothness data from \citet{Peth2016} (see Table 1). In total, we use 15 photometric features and 4 morphological features when training and testing our SOMs. Half-light radii come from a single-S\`{e}rsic fit to sources extracted from H-band images. \citet{Peth2016} extract morphological quantities from the {\sc wfc3} F125W and F160W images obtained by CANDELS. We use the H-band morphologies from the \citet{Peth2016} catalog. Training data consists of the colors (\citealp{Guo2013}) and sizes/morphologies (\citealp{Peth2016, Haussler2013}) for $\sim$500 galaxies with known spectroscopic redshifts. We match the size/morphology data to the photometry for each of the sources in these catalogs based on sky coordinates.

\begin{table}[]
\begin{tabular}{lllll}
  \cline{1-1}
  \hline
\multicolumn{1}{}{} \quad\quad\quad Feature & Wavelength ($\mu$m) &  Refs.       &  \\ \cline{1-1}
\hline\hline

 VLT/VIMOS U          & $\sim$0.36   & N09,G13 &  \\
 HST/ACS F435W        & 0.4320  & G04,K11,G13 &  \\
 HST/ACS F606W        & 0.5956  & G04,K11,G13 &  \\
 HST/ACS F775W        & 0.7760  & G04,K11,G13 &  \\
 HST/ACS F814W        & 0.8353  & G04,K11,G13 &  \\
 HST/ACS F850LP       & 0.8320  & G04,K11,G13 &  \\
 HST/WFC3 F098M       & 0.985   & W11,G13 &  \\
 HST/WFC3 F105W       & 1.045  & K11,G13 &  \\
 HST/WFC3 F125W       & 1.250  & K11,G13 &  \\
 HST/WFC3 F160W       & 1.545  & K11,G13 &  \\
 VLT/ISAAC Ks         & 2.16  & R10,G13 &  \\
 Spitzer/IRAC 3.6     & 3.6 & A13,G13 &  \\
 Spitzer/IRAC 4.5     & 4.5 & A13,G13 &  \\
 Spitzer/IRAC 5.8     & 5.8 & G13 &  \\
 Spitzer/IRAC 8.0     & 8.0 & G13 &  \\
 \hline
 R$_{50}$              & 0.4320  & H13 &  \\
 Concentration (C)    & 1.250  & P16 &  \\
 Asymmetry (A)        & 1.250  & P16 &  \\
 Smoothness (S)       & 1.250  & P16 &  \\
 
\hline
\end{tabular}
\label{table:data}
\caption{The 19 features used in the training and testing of the SOMs. The first 15 lines of the table are the photometry, showing the instrument and filter used as well as the central wavelength of the filter. The bottom 4 lines of the table show the morphological quantities used and the corresponding wavelengths. References: G04:\citet{Giavalisco2004}, N09:\citet{Nonino2009}, R10:\citet{Retzlaff2010}, K11:\citet{Koekemoer2011}, W11:\citet{Windhorst2011}, A13:\citet{Ashby2013}, G13:\citet{Guo2013}, H13:\citet{Haussler2013}, P16:\citet{Peth2016} }
\end{table}


Galaxy morphologies are captured by a number of quantities; for example, radius, concentration, asymmetry, smoothness, S\`ersic index, axis ratio, Gini coefficient, and second order moment (e.g., \citealp{Conselice2000, Conselice2003, Lotz2004, Peth2016}). A galaxy's spatial extent can be characterized through measurements of half-light radius (hereafter R$_{50}$), which is the radius at which 50\% of the galaxy's total flux falls. Concentration \citep{Kent1985, Bershady2000, Conselice2003} describes the extent to which a galaxy's light is concentrated toward the center. The concentration is taken to be the ratio between the radii containing 80\% and 20\% of the galaxy's light within 1.5 Petrosian \citep{Petrosian1976} radii (e.g., \citealp{Peth2016}). Large scale asymmetries in the light distribution of the source are described by the asymmetry statistic \citep{Conselice2000}. High asymmetry is typical for blue, star forming galaxies and can be indicative of systems that have undergone mergers \citep{Conselice2000, Conselice2003}. Smoothness \citep{Conselice2003}, also known as clumpiness, traces structures with high spatial frequencies, such as star forming regions. In contrast, objects like elliptical galaxies consist primarily of low spatial frequencies, due to their smooth light distributions. \citet{Conselice2003} define clumpiness as the ratio between the flux in high frequency spatial structures and the total flux of the galaxy. There are alternative methods for identifying clumps, such as resolved rest-frame (U-V) color selections (\citealp{Hemmati2014}, see also \citealp{Wuyts2012,Guo2015}) which yield comparable results.

Together, concentration, asymmetry, and smoothness make the CAS structural parameter system \citep{Conselice2003}. The CAS parameters form a three-dimensional volume that can be used to classify galaxies into elliptical, spiral, dwarf irregular, dwarf elliptical and merger classes. We include the CAS system in our analysis to see if the evolution of morphological parameters correlates strongly enough with redshift in order to improve photo-$z$ estimates.

We provide a brief summary of other interesting morphological quantities that could also potentially be used in training the self-organizing maps, though were not used in this study. The Gini coefficient \citep{Lorenz1905, Abraham2003, Lotz2004} is a quantity used to measure how equally light is distributed amongst pixels in a galaxy image. The Gini coefficient is also correlated with concentration \citet{Abraham2003}. The second-order moment \citet{Lotz2004} measures the flux in pixels weighted by their squared distance from the galaxy center. This statistic is sensitive to bright features like galactic nuclei, bars, spiral arms, and star clusters \citet{Lotz2004}.


\section{Redshift Measurement Algorithm}

We use the self-organizing map to identify correlations between redshift and observed galaxy colors as measured from the multi-band optical and near-infrared data. Galaxy morphological information is included in the self-organizing map algorithm in a later section. When the SOM is given the color/morphology data of a test galaxy, it searches for the node that is closest in color-morphology space to that test galaxy and makes an approximation of its redshift based on the location of the node within the map. In theory, we could supply the self-organizing map with any observable quantity (photometric or morphological; such as color, half-light radius, S\`ersic index, asymmetry, concentration, Gini coefficient, etc.), and the SOM would cluster the input data according to the correlations that it locates in the data. For galaxy SED studies, this means that we can explore any of the mapped properties and associate those with a measured value given the clustered information.

The construction of the self-organizing map is similar to the self-organizing map association network (SOMA) from \citet{Yamakawa2001}, though our method of association differs. A SOMA infers a set of perfect (complete) information from a set of incomplete information. For the case presented here, we take the perfect information to be a vector of data points consisting of galaxy photometry, morphology, and spectroscopic redshift, and the incomplete information would be a vector of photometric and morphological data points, without a redshift. The SOMs are constructed and organized from a set of training samples consisting of perfect information; subsequently, samples composed of incomplete information and unknown spectroscopic redshift can be presented to the map for redshift classification. Note that {\it perfect} in this sense does not mean without error, but rather that the data {\it exists}.

The self organizing map is initialized to an $m\times n$ array of nodes. Each node contains a weight vector that covers the attribute (e.g., color, size, spectroscopic redshift) space of the input data. This weight vector is intialized to random values, and, as the map is trained, these weight vectors will update themselves to be more representative of the data. This training process is repeated for each galaxy in the training sample. The map as a whole has a topology which we take to be toroidal. Various works in the literature (e.g., \citealp{Yamakawa2001, Masters2015}) describe the training process in detail. We summarize the same process here and borrow their notation. One training iteration begins with the selection of a random training sample with feature vector $\vec{x}$ containing photometric and morphological data as well as a spectroscopic redshift. Next is the identification of the Best-Matching Unit (BMU), the node which is closest in attribute space to the training sample according to the reduced-$\chi^2$ distance given by

\begin{equation}
  d^2_k(\vec{x}, \, \vec{w_k}) = \frac{1}{m}\sum_{i=1}^{m} \frac{(x_i - w_{k,i})^2}{\sigma^2_{x_i}}
  \label{equation:chisquared}
\end{equation}

where $d_k$ is the reduced-$\chi^2$ distance, m is the length of the feature vector $\vec{x}$, $x_i$ is the $i^{th}$ component of $\vec{x}$, $\sigma_{x_i}$ is the uncertainty associated with $x_i$, and $\vec{w_k}$ is the $k^{th}$ weight vector in the SOM. In the cases in which a training object or testing object was missing a data feature (i.e., a value of -99 for flux in some band), the reduced $\chi^2$ distances for each node were computed by taking the missing feature to be exactly equal to the node weight that corresponded to the missing feature; i.e., setting $x_i$ equal to $w_{k,i}$ for that feature. This means that only the non-missing data will contribute to the sum in Equation \ref{equation:chisquared}. In this way, the incomplete training/testing vector can still exist in the $m$-dimensional feature space, but its reduced $\chi^2$ distance will only depend on the features that are not missing. This technique also works if more than one feature are missing.

The goal is to have nodes with similar weights located near each other in the map. The nodes in the ``neighborhood'' of the BMU are determined by the neighborhood function $H_{k}$, which we take to be Gaussian:

\begin{equation}
  H_{k}(t) = e^{-d^2_{k}/\sigma^2(t)}
\end{equation}

where the standard deviation $\sigma^2(t)$ of the neighborhood function is:

\begin{equation}
  \sigma(t) = \sigma_0 \bigg(\frac{1}{\sigma_0}\bigg)^{(t/N_{iters})}
\end{equation}

where $\sigma_0$ is an arbitrary initial value, and $t$ is an integer ranging from 1 to the total number of training iterations, $N_{\rm{iters}}$.

The BMU and surrounding nodes are then rewarded for being nearest to the training sample and are allowed to update their weights according to the relation:

\begin{equation}
  \vec{w_k}(t+1) = \vec{w_k}(t) + a(t) H_{k}(t)[\vec{x}(t) = \vec{w_k}(t)]
  \end{equation}

where we adopt the learning function $a(t)$:

\begin{equation}
  a(t) = e^{-(t/N_{iters})}
\end{equation}

While other learning functions exist in the literature (e.g., \citealp{Masters2015}), we selected this one because it gave the lowest outlier fraction.
The learning function decreases monotonically and is intended to de-sensitize the SOM to new training data as time progresses, allowing it to converge to a stable solution.


The multitude of SOM parameters (e.g., number of nodes, number of training iterations, learning rate, neighborhood function) affect the performance of the SOM as a whole. The number of nodes and training iterations used will depend on the total number of training samples available. A larger training set will require more training iterations to fully capture the data; however, it is possible to over-train a map with too many training iterations, where the SOM learns the training data well but does not generalize to data it has not seen before.
The number of nodes affects the number and size of clusters that form in the trained map. If the number of nodes is too small, the map may not capture the full set of relations present in the data. Increasing the number of nodes and training iterations comes at a cost in computing time as well. We determined by cross-validation that a map size of 150 pixels by 150 pixels had optimal predictive ability. Cross-validation involves removing a subset of samples (the validation set) from the training set, training the map on the remaining samples, and then using the validation set as testing samples. The grid size of the map is varied, and the optimal value of the grid size hyperparameter is selected based on performance on the validation set.

To extract a redshift prediction from the SOM, it is presented with a test vector that contains the same photometric and morphological attributes as the training vectors, but without the spectroscopic redshift. While ignoring the redshift attribute of the SOM nodes, the reduced-$\chi^2$ distance is computed between the test vector and each node in the map, identifying the best-matching unit (node). The redshift of the best-matching unit becomes the redshift associated with the test vector and represents the best prediction of the redshift of the test source.


\section{SOMs on Galaxies}

In order to test the SOM, the known spectroscopic redshifts of galaxies must be compared to the predictions of the map. However, the galaxies used to test the map must not be sources that the map has seen before; that is, they cannot appear in the training set. A study of several photometric redshift codes was performed by \citet{Dahlen2013}, and they have released the training and control catalogs based on GOODS-S data that was used in the study. As a first test, our SOMs were trained and tested using this training/control set, which contained only photometric data. For each source, the quantity $\sigma = \Delta z / (1 + z_{\rm spec})$, where $\Delta z = z_{\rm BMU} - z_{\rm spec}$, is determined. There are several measures of performance (e.g., \citealp{Dahlen2013}), denoted by $\sigma_F$ ($= rms[\Delta z / (1 + z_{\rm spec})]$), $\sigma_O$ (the same as $\sigma_F$ but has sources with $\sigma \, > \, 0.15$ removed), and the outlier fraction (OLF) specifying the fraction of sources with $\sigma \, > \, 0.15$). Individual SOMs were trained using the training/testing set from \citet{Dahlen2013}, and the performance of individual maps was found to be $\sigma_F\sim 0.17$, $\sigma_O\sim 0.042 - 0.044$, and OLF $\sim 9\% - 10\%$. To obtain a slight improvement in accuracy, the median of the results of 500 SOMs was found (since each self-organizing map will be slightly different as the initial node weights are random and the training samples may be presented in different orders), giving $\sigma_F \sim 0.15$, $\sigma_O \sim 0.036 - 0.038$, and OLF $\sim 6\% - 8\%$.

\begin{figure*}[!th]
  \centering
    \includegraphics[trim=0cm 3cm 0cm 0cm, scale=0.8]{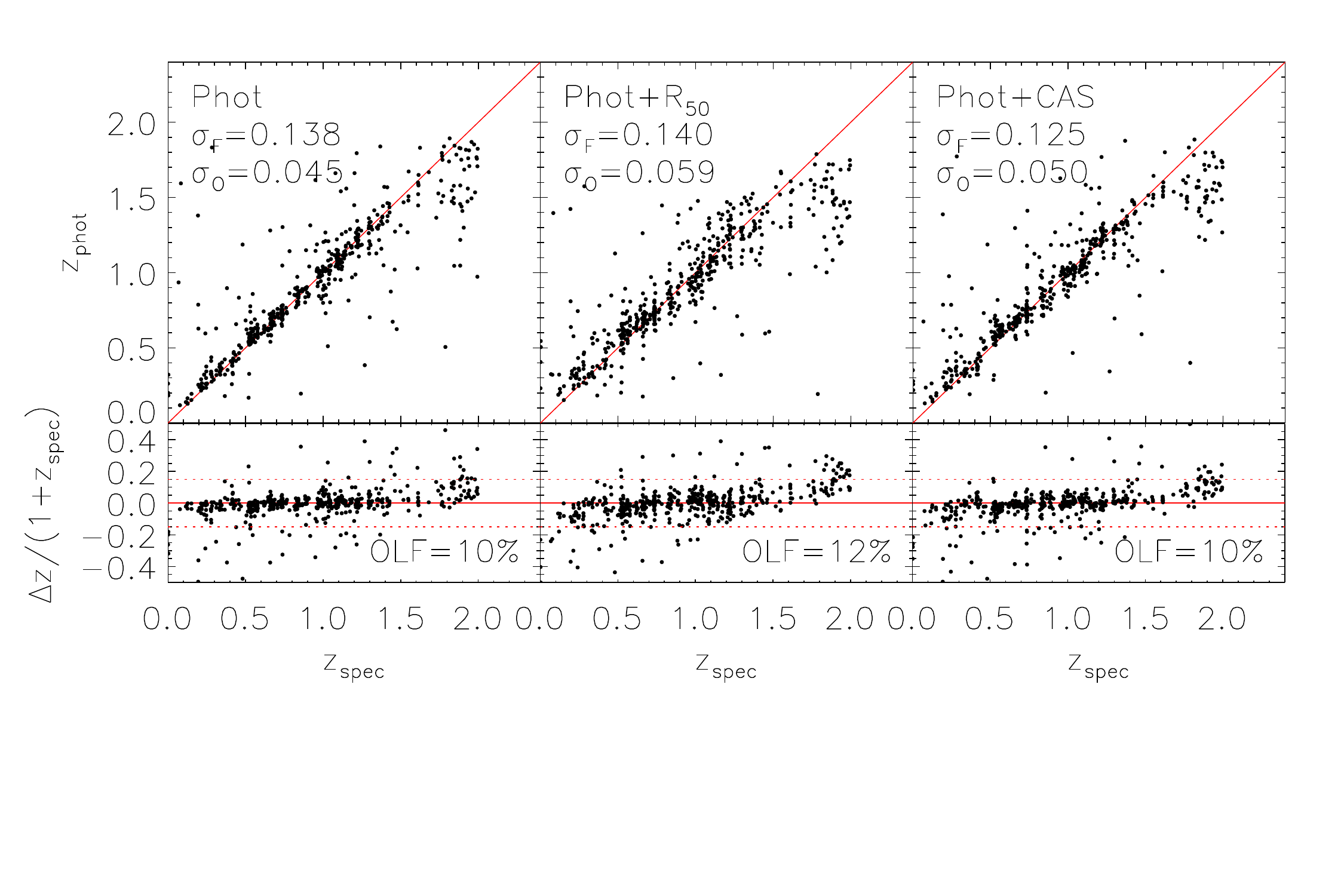}
  \caption{The top row shows a comparison of photo-z to spec-z for GOODS-S field using different subsets of data features. The bottom shows the normalized residuals given by ($z_{\rm phot} - z_{\rm spec}$)/(1 + $z_{\rm spec}$). Left: SOM predictions using only photometric data. Middle: Using photometry and half-light radius. Right: Using photometry and concentration-asymmetry-smoothness (CAS) data.}
\label{fig:redshift_comparison}
\end{figure*}

Next, we trained and tested the SOMs using three training/testing set pairs each composed of $\sim$500 sources with $z < 2$. The first training/testing set contained only 13 colors (computed from 14 photometric bands), the second set contains R$_{50}$ from a single-S\`{e}rsic fit in addition to the colors, and the third set contains the colors as well as concentration, asymmetry, and clumpiness (CAS) data. We select sources with $z \, < \, 2$ because morphological measurements for higher redshift sources will be inherently less precise. A single self-organizing map trained and tested with our training set of $z < 2$ sources produced a typical $\sigma_F$ in the range 0.14 - 0.16 and $\sigma_o$ in the range 0.048 - 0.052 with outlier fractions of $\sim$ 10\% - 12\%. By computing the median of multiple SOMs, we produced slightly lower values of $\sigma_o$. By averaging the SOM outputs in this way, we obtained the results in Table 2 when using photometry alone, and photometry with either $R_{50}$ or CAS. An example of typical results is shown in Figure \ref{fig:redshift_comparison}.

\begin{table}[]
\begin{tabular}{lllll}
  \cline{1-1}
  \hline
\multicolumn{1}{}{}    & $\sigma_{F}$  & $\sigma_{o}$   & OLF$^{1}$         &  \\ \cline{1-1}
\hline\hline
photometry only        & $\sim$0.14   & $\sim$0.05 & 10\% - 11\%              &  \\
with R$_{50}$           & $\sim$0.14   & $\sim$0.06 & 12\% - 14\%       &  \\
with CAS               & $\sim$0.13   & $\sim$0.05 & 10\% - 12\%       &  \\
\hline
\end{tabular}
\label{table:results}
\caption{Summary of performance when using the median of multiple SOM predictions after training was done with photometry alone, photometry plus half-light radius, and photometry plus concentration, asymmetry, and smoothness. The addition of morphological parameters had an insignificant effect on photometric redshift estimation. $^{1}$OLF: Outlier fraction, the fraction of sources with $\sigma \, > \, 0.15$.}
\end{table}

For comparison, we run several public photo-$z$ codes on the three training/testing set pairs. The photo-$z$ codes used were {\sc PhotoRApToR} using {\sc mlpqna} (\citealp{Brescia2013}), {\sc flexcode} (\citealp{Izbicki2017}), and {\sc tpz} and {\sc somz} from the {\sc mlz} package (\citealp{CarrascoKindBrunner2013, CarrascoKindBrunner2014}). Here we will only give a brief summary of these algorithms.
{\sc mlpqna} uses a supervised learning technique involving multi-layer perceptrons, a network of neurons that is trained by minimizing a loss function. The loss function is minimized by iteratively updating the weights in the neural network. The Quasi-Newton Algorithm is used to compute the Hessian of second derivatives, which is necessary for computing the amount by which the network weights are updated. We use a three layer network with 15, 16, or 18 neurons in the first layer (if the training set contains just photometry, phot + R$_{50}$, or phot + CAS, respectively), 64 neuron in the second layer, and 1 neuron in the final layer. We set a decay rate of 0.001 and use 10000 max iterations.

{\sc flexcode} employs a conditional density estimator method which seeks to improve photo-zs by constructing a full conditional density distribution from the data. This is done using an orthogonal series formulation, with the series coefficients determined by regression. The result is a conditional probability distribution that is useful for handling the multi-modality in a photo-z prediction. When running {\sc flexcode}, we use the XGBoost regression method with a cosine basis system.

{\sc mlz} can perform regression using two different methods: a prediction tree and random forest algorithm and a self-organizing map algorithm. Prediction trees work by splitting the data into multiple branches based on some attribute. This process is repeated recursively until a stopping criterion is met, at which point a photo-z prediction can be made. A random forest is a collection of prediction trees whose predictions can be combined to produce more accurate results. The SOM component of {\sc mlz} works similarly to the SOM algorithm described in this work. The main difference between the SOM algorithms is the way in which spectroscopic redshift is used to train the SOM. In the {\sc mlz} {\sc somz}, the spectroscopic redshift does not enter in the training of the SOM. Only after the map has been trained are the spectroscopic redshifts from the training sample associated with the nodes in the map, with the mean redshift of the sources associated with each node becoming the final redshift of that node. For our study with {\sc tpz}, we set the MinLeaf parameter to 10. For {\sc somz}, we use a periodic grid with a size of 64 nodes and 3000 training iterations.

\begin{figure*}[!th]
  \centering
    \includegraphics[trim=0cm 3cm 0cm 0cm, scale=0.8]{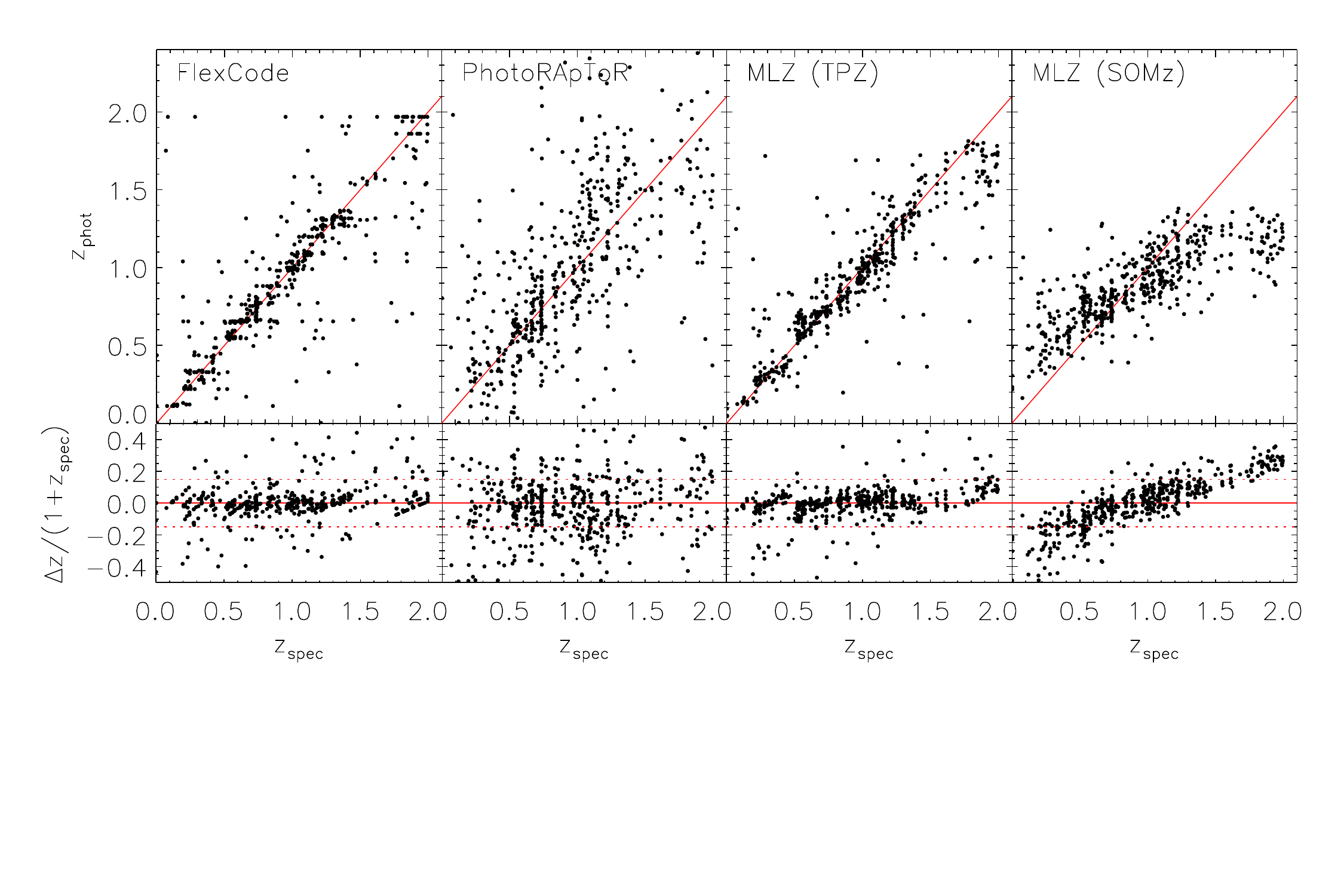}
  \caption{An example of the results from the literature photo-z codes when applied to our training/testing set containing photometry and R$_{50}$.  See text for references and Table 3 for quantitative metrics of the results. We find that our SOM implementation produces results that are similar in dispersion and outlier fraction. PhotoRApToR and SOMz had unusually large outlier fractions, which we attribute to the effects of missing data in the training/testing sets. It is possible that a more extensive search over hyperparameter space may yield better results. }
\label{fig:photoz_software_comparison}
\end{figure*}

Our implementation of the SOM algorithm uses a supervised approach. The spectroscopic redshift is included during the training process, and the final trained map will contain weights corresponding to the final redshift associated with each node. Overall, the performances of our SOM algorithm and the other photo-z codes were comparable, though missing data negatively affected the performance of some of the codes. As almost every source was missing photometry in one band or another, the replacement of the missing value with -99 may not allow the codes to perform optimally, while at the same time, removal of all data points with a missing value was not possible. The results from the photo-z codes are shown in Figure \ref{fig:photoz_software_comparison}, and the corresponding metrics are listed in Table 3. {\sc flexcode} returned similar results for all three testing sets. The {\sc tpz} algorithm from {\sc mlz} was generally less accurate for the testing sets that included morphological data. We note that it is possible that there may exist hyperparameters for the {\sc flexcode} and {\sc tpz} algorithms that may improve their predictions but which we may have missed while tuning these models, despite our best efforts to find the optimal hyperparameters. {\sc mlpqna} and the {\sc somz} algorithm had large outlier fractions, with the number of outliers increasing when morphological data was used in training. It is likely that the large outlier fractions may be caused by missing data.

\begin{table}[]
\begin{tabular}{lllll}
  \cline{1-1}
  \hline
\multicolumn{1}{}{}    & $\sigma_{F}$  & $\sigma_{o}$   & OLF         &  \\ \cline{1-1}
\hline\hline
FlexCode                & $\sim$0.15   & $\sim$0.05  & 11\% - 13\%              &  \\
PhotoRApToR (MLPQNA)    & $\sim$0.44   & $\sim$0.07  & 21\% - 27\%            &  \\
MLZ (TPZ)               & $\sim$0.12   & $\sim$0.05  & 9\% - 10\%             &  \\
MLZ (SOM)               & $\sim$0.16   & $\sim$0.07  & 24\% - 28\%            &  \\
\hline
\end{tabular}
\label{table:photz_software_results}
\caption{Typical results obtained by running photo-z codes from the literature on our training/testing sets including photometry and morphologies. The results from our SOM implementation are about the same as the results from these other softwares.}
\end{table}

\section{Probability Distributions}

\begin{figure}
  \includegraphics[scale=0.43, trim=0cm 0cm 0cm 0cm]{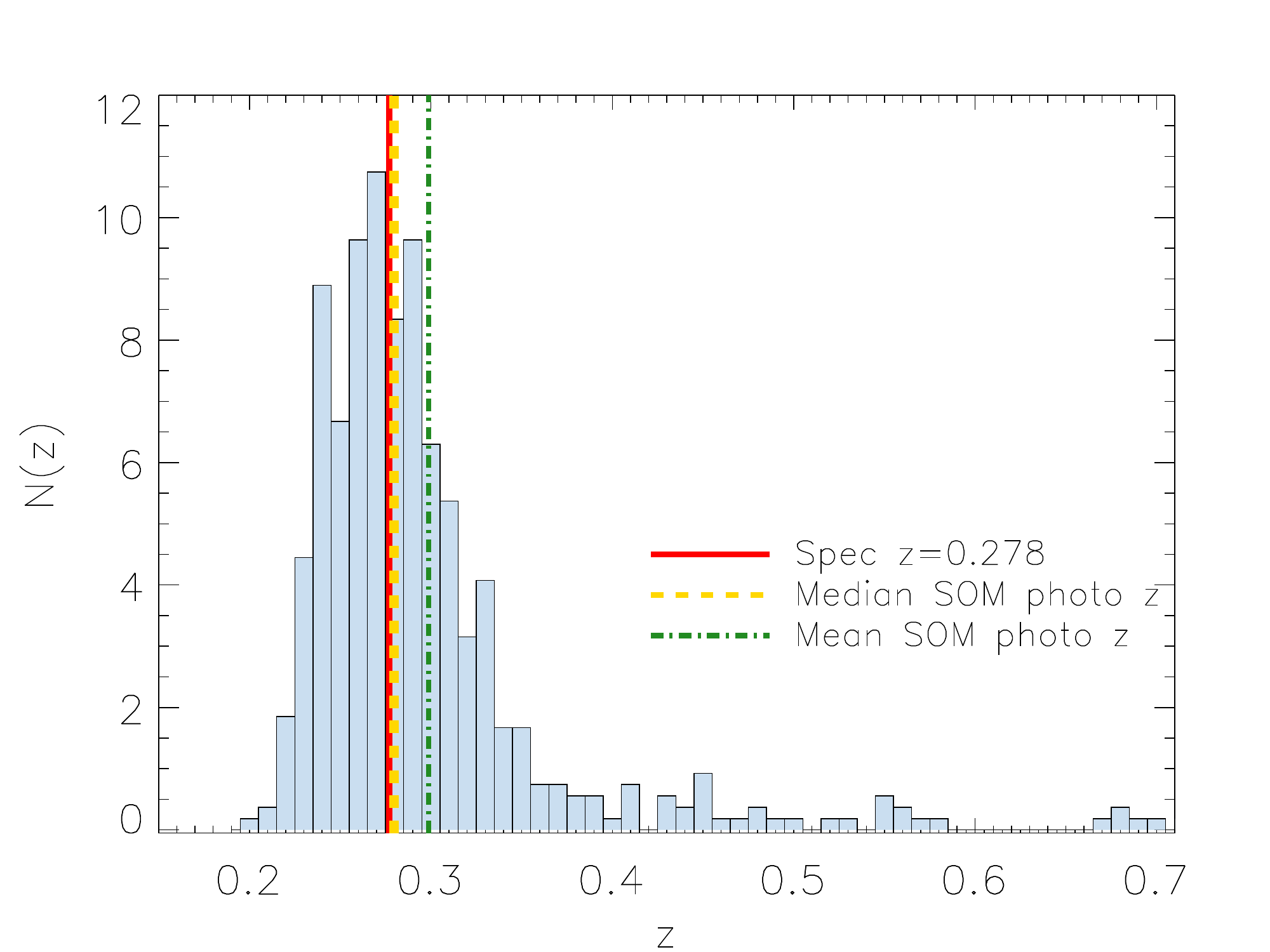}
\caption{Example of a redshift probability distribution generated using 500 different SOMs. The spectroscopic redshift for this source is $z = 0.278$. Since each of the 500 SOMs is initialized with a different random set of parameters, each will converge to its own estimate of the redshift. The median of multiple SOMs provided measurements that were more closely aligned with the spectroscopic redshifts, due to its insensitivity to outliers.}
\label{fig:probability}
\end{figure}

Many photo-$z$ methods return a probability distribution in redshift space (e.g., {\sc lephare}: \citealp{Arnouts1999, Ilbert2006}, {\sc probwts}: \citealp{Cunha2009}) as methods that only give point estimates can miss important information; e.g., a probability distribution may be double-peaked, but a point estimate may only see the larger peak and miss the information in the secondary peak (\citealp{Mandelbaum2008, Cunha2009, Wittman2009, Bordoloi2010, Abrahamse2011, Sheldon2012}). By using an ensemble of SOMs, the algorithm that we employ can be extended to return a probability distribution. Each individual SOM in the ensemble is initialized randomly, with no two SOMs having the same starting parameters. The different initializations will lead each map to converge to different weights after the training process is completed, and thus each map will predict a different photometric redshift for a test source. The results from the ensemble of SOMs are histogrammed with a bin size of $\Delta z$ = 0.01 to form the final probability distribution function (see Figure \ref{fig:probability}), and the median of the distribution is taken to be the final point estimate of the redshift.

\begin{figure}[!th]
  \centering
    \includegraphics[trim=0cm 0cm 0cm 0cm, scale=0.4]{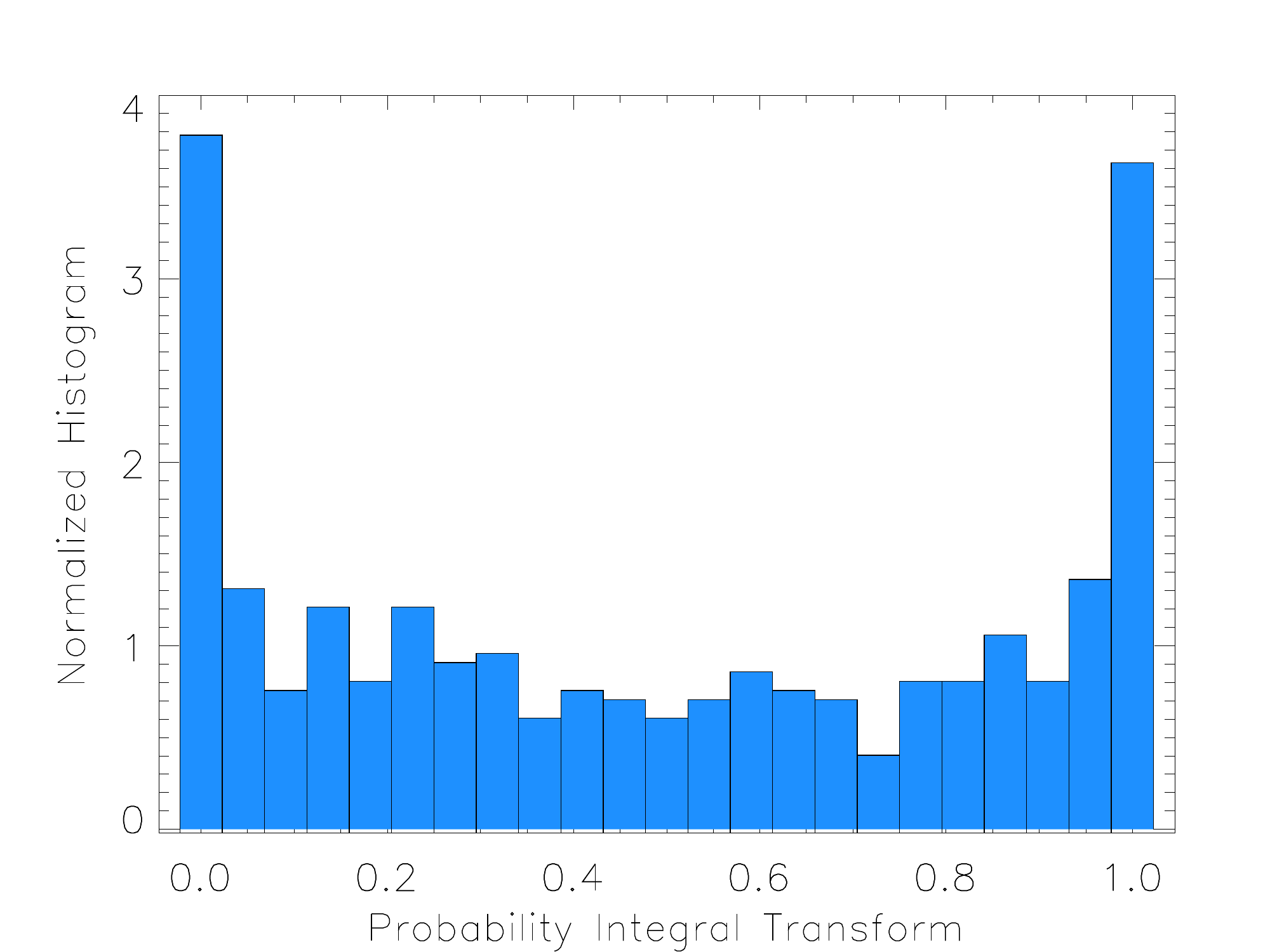}
  \caption{The probability integral transform (e.g., \citealp{Polsterer2016}) from a set of redshift probability distribution functions. A set of well-calibrated PDFs will have a near uniform PIT. The U-shape of our PIT indicates that our redshift PDFs are underdispersed.}
\label{fig:pit}
\end{figure}

The quality of the probability distribution functions (PDFs) is tested using the probability integral transform (PIT) described in \citet{Polsterer2016} and the confidence test from \citet{Wittman2016}. The probability integral transform (PIT, \citealp{Dawid1984}) is given by the histogram of the cumulative probabilities of each redshift PDF computed at the value of the spectroscopic redshift. The PIT histogram serves as a visual guide for how well-calibrated the probability distribution is (\citealp{Polsterer2016}). Figure \ref{fig:pit} shows an example derived from the SOM distribution functions. Ideally, the PIT should be nearly uniform if the PDFs are well-calibrated. The U-shape of the histogram in \ref{fig:pit} indicates that our PDFs are underdispersed, i.e., that the dispersion in the redshift PDFs predicted by the SOMs is too small and the spectroscopic redshifts are too often ending up in the tails of the PDFs. As such, it appears that there is an overabundance of PDFs in which the statistical likelihood is very low for the spectroscopic redshift associated with the galaxy for that PDF. This means that the PDFs do not adequately represent the spectroscopic redshifts, and more work is required to make them more accurate.

\begin{figure}[!th]
  \centering
    \includegraphics[trim=1cm 0cm 0cm 0cm, scale=0.45]{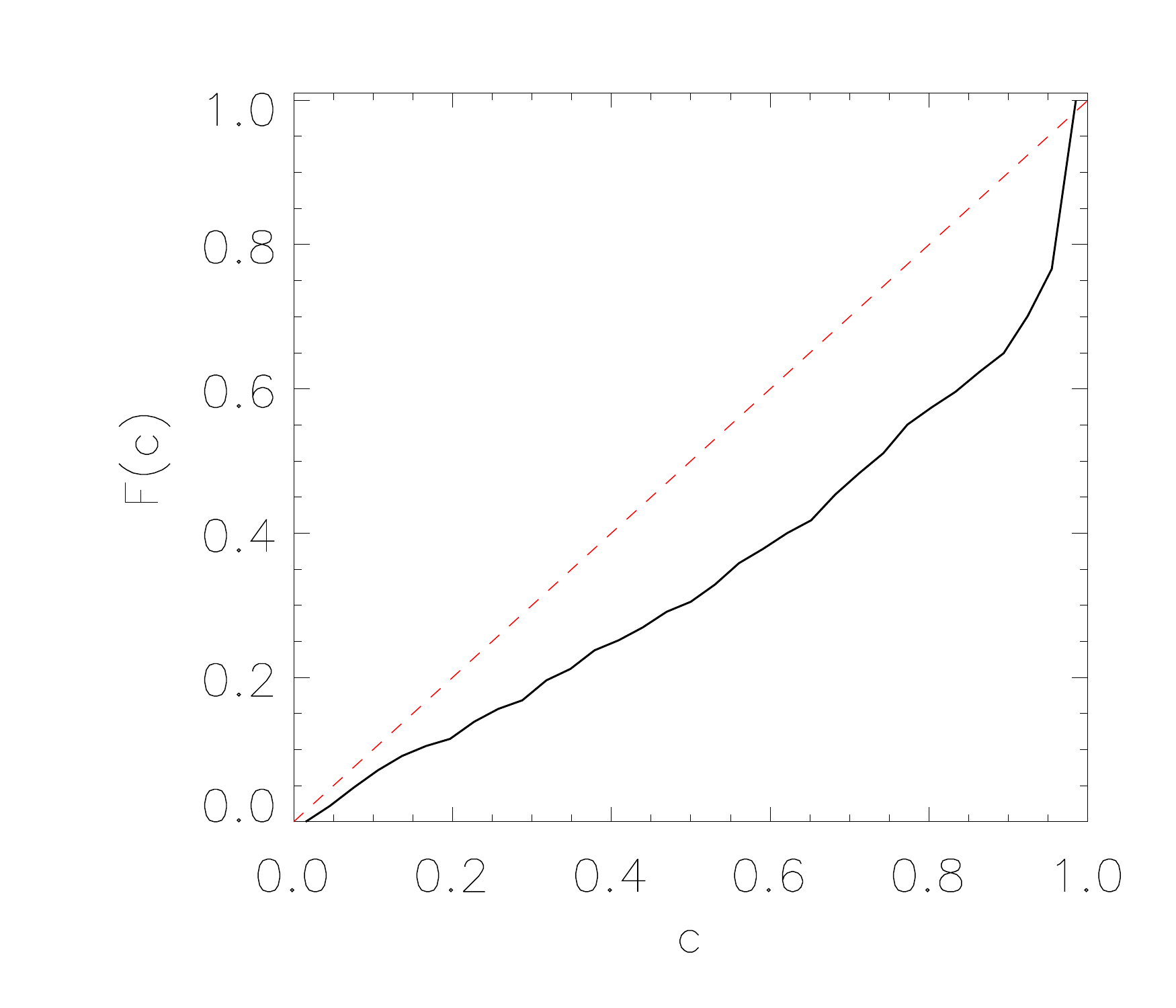}
  \caption{The confidence test from \citet{Wittman2016}. Shown in black is the cumulative distribution function, $F(c)$, of the binned threshold credibilities, $c$. The red dashed line represents the case in which the redshift probability distribution functions have a well-calibrated width. The plot indicates that at least some of our redshift PDFs are overconfident, i.e., that their widths are too narrow.}
\label{fig:qq_plot}
\end{figure}

The second metric used to test the SOM PDFs is the test developed by \citet{Wittman2016} to determine whether the widths of probability distribution functions are over- or under-confident. We refer readers to the original paper for a more in-depth explanation of the test but provide a brief summary here. This confidence test is based on the principle that, ideally, a sample of galaxies should have 1\% of its spectroscopic redshifts fall in the 1\% credibility intervals (CI) of the corresponding PDFs, 2\% of spectroscopic redshifts fall in the 2\% CI, 50\% of spectroscopic redshifts fall in the 50\% CI, and so on. To perform the test, the threshold credibility, $c_i$, is computed for each galaxy in the testing set. The cumulative probability function $F(c)$ is then found from the distribution of the $c_i$. This cumulative distribution function is plotted in Figure \ref{fig:qq_plot}. Ideally, the curve should lie on the red dashed line, if 1\% of $z_{spec}$ fall in the 1\% CI, etc. In our case, the black curve lies below the ideal case, indicating that our redshift PDFs are overconfident, i.e., that the confidence intervals are too narrow and the uncertainties are underestimated. Again, more work is needed to improve the PDFs.

\section{Discussion}

\begin{figure}[!th]
  \centering
    \includegraphics[trim=0cm 0cm 0cm 0cm, scale=0.45]{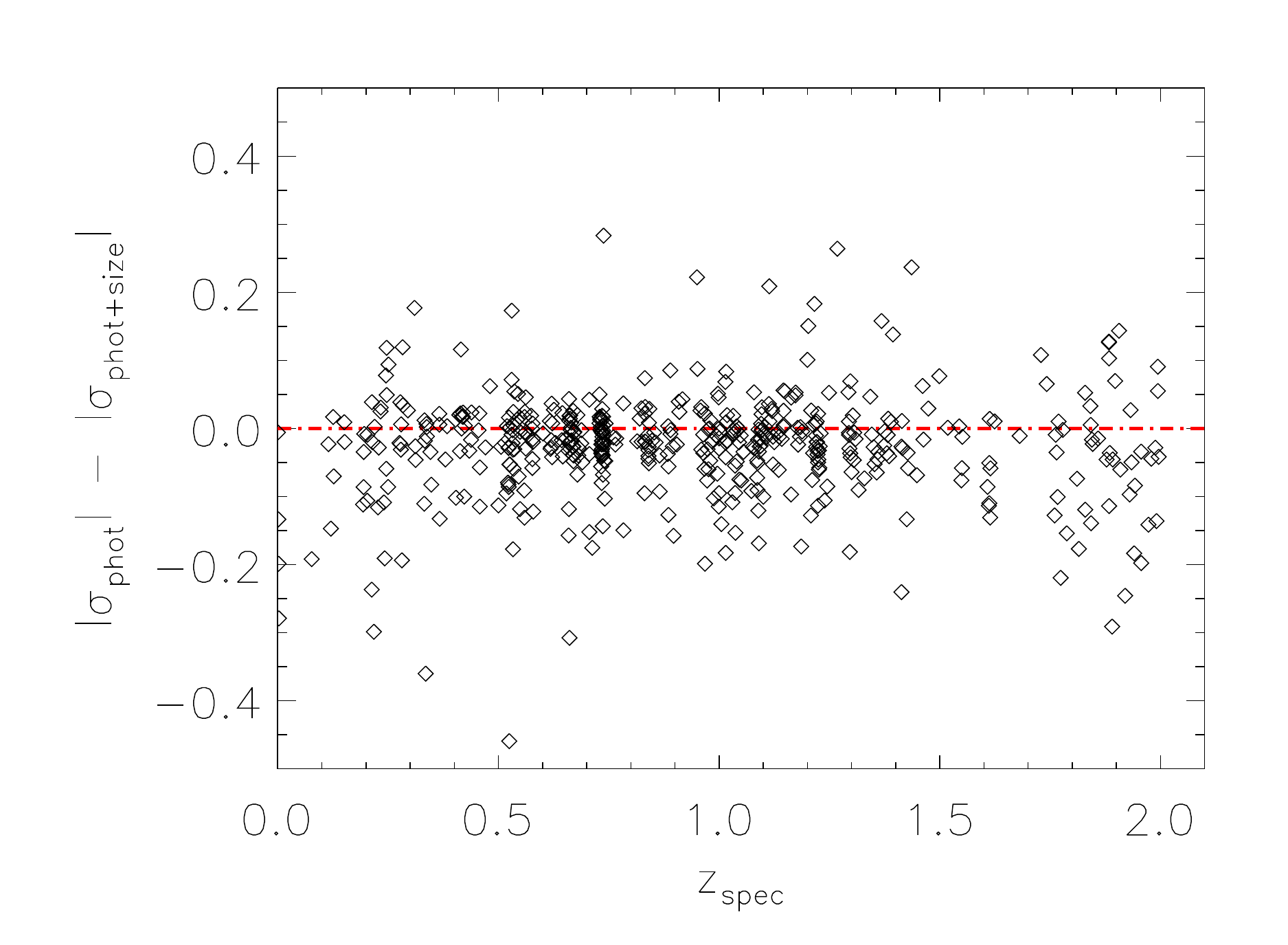}
  \caption{Comparison of SOM predictions for each test source with and without half-light radius. The quantity $\Delta$z is given by z$_{\rm{spec}}$ - z$_{\rm{phot}}$. We show $\Delta$z calculated with photometry alone ($\Delta$z$_{\rm{phot}}$) minus $\Delta$z calculated with photometry and size ($\Delta$z$_{\rm{phot+size}}$). Positive values indicate that use of half-light radius increased the accuracy of the SOM, while negative values indicate a decrease in accuracy.}
\label{fig:sigma_z}
\end{figure}

\begin{figure}[!th]
  \centering
    \includegraphics[trim=2cm 0cm 0cm 0cm, scale=0.6]{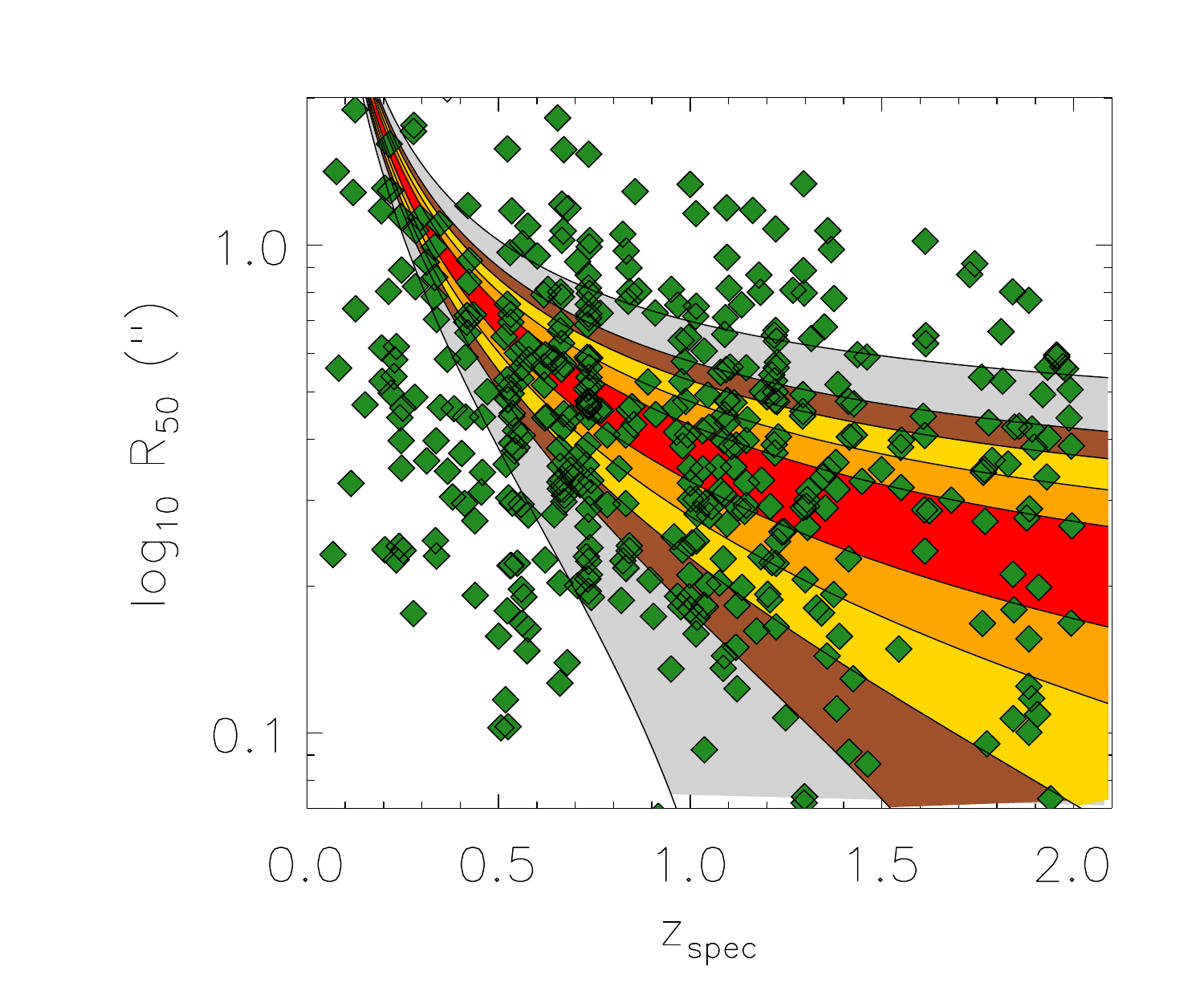}
  \caption{Comparison of simulated R$_{50}$ with real R$_{50}$ data (green diamonds). The regions correspond to simulated R$_{50}$ with different Gaussian spreads around a presumed average trend; red: $\sigma = 0.05$ arcseconds, orange: $\sigma = 0.1$ arcseconds, yellow: $\sigma = 0.15$ arcseconds, brown: $\sigma = 0.2$ arcseconds, and gray: $\sigma = 0.32$ arcseconds. The scatter of the real R$_{50}$ is $\sim$ 0.32 arcseconds, with approximately 68$\%$ of data points falling within the gray region. We find improvement in photo-z estimates that include R$_{50}$ only when the spread in R$_{50}$ is smaller than 0.05 arcseconds. Such a spread in real data may be impossible to achieve due to the intrinsic variation in R$_{50}$, even with increased telescopic precision.}
\label{fig:real_vs_sim_R}
\end{figure}

\begin{figure}[!th]
  \centering
    \includegraphics[trim=0cm 0cm 0cm 0cm, scale=0.5]{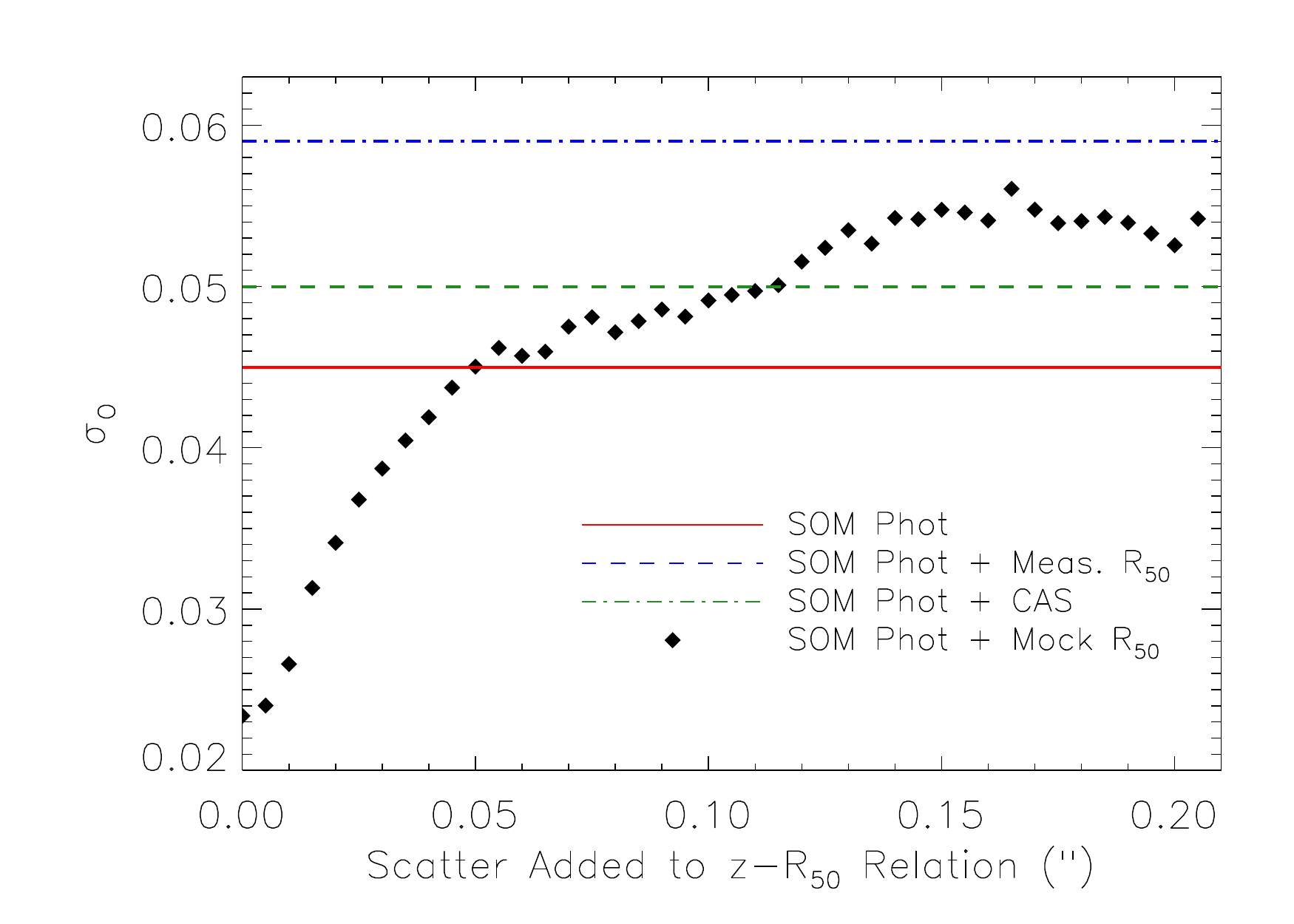}
  \caption{Redshift uncertainty as a function of the scatter added to the theoretical size relation for the GOODS-S field (black dots). The training data for the SOM results given by the black dots consist of photometry and size (half-light radius, computed according to the relation in \citet{Mosleh2012} (see also \citealp{Vanderwel2014}). For comparison, we show the performance of the SOM when using photometry alone (red line), photometry and half-light radius from GALFIT (blue line), and the existing precision of photo-zs in the CANDELS catalog. The SOMs with photometry+size would perform better than with photometry alone if the variation in size at a particular redshift was less than about 0$^{\prime\prime}$.02. If future surveys with higher precision instruments could measure half-light radii to this precision, the SOM networks presented here may offer improvement to photo-z estimates.}
\label{fig:scatter}
\end{figure}

Figure \ref{fig:sigma_z} shows the difference between the SOM photo-z using photometry alone and the SOM photo-z using photometry in conjunction with R$_{50}$. For each galaxy in the test sample, we calculate its redshift with and without R$_{50}$ as input data and then determine the absolute difference between the two photo-zs ($|\Delta z_{\rm phot}|$ and $|\Delta z_{\rm phot+size}|$) and the spectroscopic redshift. If R$_{50}$ had no effect on the redshift determination, then $|\Delta z_{\rm phot}| - |\Delta z_{\rm phot+size}|$ should be zero. If, however, R$_{50}$ led to some improvement, then $|\Delta z_{\rm phot}| - |\Delta z_{\rm phot+size}|$ would be positive, since the deviation of z$_{phot}$ from z$_{spec}$ would be larger than the deviation of $z_{\rm phot+size}$ from $z_{\rm spec}$. Negative values would indicate that R$_{50}$ had a detrimental effect. In Figure \ref{fig:sigma_z}, 67\% of data points lie below zero, indicating that half-light radius did not improve photo-z estimation.

We find that the addition of galaxy morphological data does not significantly improve the redshift estimation from the self-organizing maps. The scatter introduced by the morphological data most likely dominates any benefit coming from the correlation between redshift and morphology. These results appear to be in line with the results from \citet{Soo2018}, who find that adding morphological quantities such as galaxy size, S\`{e}rsic index, surface brightness, and ellipticity do not significantly improve photo-z estimates when combined with a complete set of good photometry (in their case, full {\it ugriz} photometry). \cite{Soo2018} conclude that including a full set of photometric bands may saturate the amount of redshift information available, which is reasonable given that they find improvement in photo-z estimates when morphology is used in conjunction with sub-optimal photometry or photometry in fewer than all five {\it ugriz} bands. Similarly, we conclude that our use of morphology, at its present precision, may not be providing any new information that is not already contained in our 15 bands of photometry. \citet{Soo2018} also compare the effects of low-quality versus high-quality morphology by studying galaxy radii measured by the Sloan Digital Sky Survey (SDSS) Stripe-82 survey and by the Canada-France-Hawaii Telescope (CFHT) in Stripe 82 (CS82), the latter of which they assume to be of higher quality due to its 0.6 arcsecond seeing. However, they do not find any improvement in photo-zs when using the CS82 data over the SDSS data. In comparison, we find that improvement might be possible if the scatter in radii is less than 0.05 arcseconds (Figure \ref{fig:scatter}), which is well below the CS82 seeing.

While morphological parameters did not lead to significant increases in accuracy, we would like to see if future morphological measurements with increased precision may lead to better SOM predictions. To do this, we pass simulated R$_{50}$ data to the SOMs during training and testing. The mock size data is generated by taking the power law fits for $log(r_e)$ as a function of redshift for Lyman-break galaxies in \cite{Mosleh2012} to be the true relation between size and redshift (see also, \citealp{Vanderwel2014}). The simulated R$_{50}$ are drawn from a Gaussian distribution with a variable standard deviation (scatter) and mean equal to the half-light radius at each redshift from the ``true relation''. Figure \ref{fig:real_vs_sim_R} shows a comparison of the simulated R$_{50}$ with the actual R$_{50}$ from the data. In Figure \ref{fig:scatter}, we examine the effect that increased precision in R$_{50}$ has on $\sigma_{o}$ for a sample of galaxies. As the amount of scatter (black points) is lowered, improvement to photo-$z$ estimation is achieved when the deviation in half-light radius from the theoretical relation is less than ~0.05 arcseconds. Even with next generation space telescopes such as the James Webb Space Telescope (JWST) and the Wide Field Infrared Survey Telescope (WFIRST) with diameters of 6.5$\,$m and 2.4$\,$m, respectively, the best angular resolution possible would be ~0$^{\prime\prime}$.05 and ~0$^{\prime\prime}$.15 for  H band at ~1.65$\,\mu$m. Improvement to photo-z estimation using half-light radius may not be viable in the near future. It may also be the case that the intrinsic scatter in radii at the same redshift may be too large (i.e., greater than 0$^{\prime\prime}$.03) for any correlation to improve redshift estimates.

\section{Summary}

We apply the self-organizing map algorithm to photometric and morphological data in the GOODS-S field to study the effect that morphological parameters have on estimating photometric redshifts. The self-organizing maps are trained on photometry in 15 wavelength bands and on half-light radius, concentration, asymmetry, and smoothness for about 500 galaxies with known spectroscopic redshifts up to $z \, \sim \, 2$. The SOMs make predictions for the redshifts of about 500 galaxies in a separate testing set and are compared to the spectroscopic redshifts of those sources. The results indicate no significant improvement in the accuracy of the SOM redshift predictions when using morphology plus photometry, in comparison to photometry alone. Similar results are obtained after cursory studies using our training and testing data on other photo-z codes, leading to typical results of $\sigma_F \sim 0.13 - 0.16$, $\sigma_O \sim 0.05 - 0.07$, and OLF $\sim 10\% - 14\%$ in the best cases. We attribute this result to the large scatter in the morphological data and the possibility that morphology is not introducing any new information that is not already contained in the photometry.

Redshift probability distribution functions are produced by the SOMs in addition to point estimates. Probability distribution functions are more sensitive to multi-modality in the SOM prediction results. At the present, tests of our redshift pdfs show that they are underdispersed as well as overconfident (or too narrow in width), and more work is required to improve their accuracy.

Lastly, we explore the effect that a strong radius-redshift relation would have on the SOM predictions. The goal was to identify how tight  a radius-redshift relation would have to be in order to give improvement in photo-z estimation. This was done by simulating half-light radii with varying levels of scatter around a theoretical radius-redshift relation. The simulated radii were used along with photometry to train and test a group of SOMs. Improvement was found only for very small scatter less than $\sim \, 0.05$ arcseconds around a theoretical radius-redshift relation.

\section{Acknowledgments}

This material is based upon work supported by the National Science Foundation under award number 1633631. Additional support for this paper was provided in part by GAANN P200A150121, NSF grant AST-1313319, NASA grant NNX16AF38G, HST-GO-13718, HST-GO-14083. This work is based on observations taken by the CANDELS Multi-Cycle Treasury Program with the NASA/ESA HST, which is operated by the Association of Universities for Research in Astronomy, Inc., under NASA contract NAS5-26555.


\bibliographystyle{apj}
\bibliography{som_photz}

\end{document}